\documentclass[prl,twocolumn,amsmath,amssymb,longbibliography,showpacs, superscriptaddress]{revtex4-2}

\usepackage{comment}
\usepackage[dvipsnames]{xcolor}
\definecolor{linkcolor}{rgb}{0.2,0.2,1.0} 
\usepackage[pdftex,colorlinks=true, linkcolor= linkcolor, citecolor= linkcolor, urlcolor= linkcolor, hyperindex=true,hyperfigures=true]{hyperref} 

\usepackage{graphicx}
\usepackage{dcolumn}
\usepackage{bm}
\usepackage{times}
\usepackage[version=4]{mhchem}

\definecolor{newcolor}{rgb}{0,0,0} 
\newcommand{\empir}[1]{{#1}_{\rm emp}}

\newcommand{\va}{\text{var}}

\begin{document}

\title{Effective estimation of entropy production with lacking data}

\author{Marco Baiesi}
 \affiliation{Department of Physics and Astronomy, University of Padova, Via Marzolo 8, I-35131 Padova, Italy.}
  \affiliation{INFN, Sezione di Padova, Via Marzolo 8, I-35131 Padova, Italy.}%
\author{Tomohiro Nishiyama}
 \affiliation{Independent Researcher, Tokyo 206–0003, Japan.}
\author{Gianmaria Falasco}%
  \email{gianmaria.falasco@unipd.it}
 \affiliation{Department of Physics and Astronomy, University of Padova, Via Marzolo 8, I-35131 Padova, Italy.}
  \affiliation{INFN, Sezione di Padova, Via Marzolo 8, I-35131 Padova, Italy.}

\date{\today}

\begin{abstract}
\textbf{Abstract. }
Observing stochastic trajectories with rare transitions between states, practically undetectable on time scales accessible to experiments, makes it impossible to directly quantify the entropy production and thus infer whether and how far systems are from equilibrium.
To solve this issue for Markovian jump dynamics, we show a lower bound that outperforms any other estimation of entropy production (including Bayesian approaches) in regimes lacking data due to the strong irreversibility of state transitions. Moreover, in the limit of complete irreversibility, our new effective version of the thermodynamic uncertainty relation sets a lower bound to entropy production that depends only on nondissipative aspects of the dynamics. Such an approach is also valuable when dealing with jump dynamics with a deterministic limit, such as irreversible chemical reactions.
\end{abstract}

\maketitle

\paragraph*{\textbf{Introduction.}} Energy transduction and information processing---the hallmarks of biological systems---can only happen in finite times at the cost of continuous dissipation.
Quantifying time irreversibility and entropy production is thus a central challenge in experiments on mesoscopic systems out of equilibrium~\cite{ciliberto2017experiments,li2019quantifying}, including living matter~\cite{martin2001comparison,battle2016broken,turlier2016equilibrium,dabelow2019irreversibility,yang2021physical,ro2022model,fodor2022irreversibility,vsr} 
and technological devices~\cite{gumucs2023calorimetry}. Experimental efforts have been paralleled by the recent development of nonequilibrium statistical mechanics, focused on adding methods for estimating the entropy production rate $\sigma$ in steady states~\cite{loos2020irreversibility,ehrich2021tightest,roldan2021quantifying,wachtel2022transduction,cates2022stochastic,dechant2021improving,vsr, man20,otsubo2020estimating,kim2020learning}. Several approaches, for example, focus on improving estimates of $\sigma$ in cases of incomplete information and coarse-graining~\cite{bilotto2021excess,dieball2022mathematical,ghosal2023entropy,nitzan2023universal,vandermeer2023time}. Sometimes, they are based on lower bounds on $\sigma$~\cite{li2019quantifying,man20,otsubo2020estimating}, of which the thermodynamic uncertainty relation (TUR) is a prominent example~\cite{bar15,macieszczak2018unified,gingrich2017inferring,hasegawa2019fluctuation,timpanaro2019thermodynamic,dechant2020,falasco2020unifying,van2020entropy,dechant2021improving,van2023thermodynamic}. Classical methods estimate $\sigma$ by adding local contributions from non-zero fluxes between states~\cite{zeraati2012entropy}. However, recent approaches have also introduced methods exploiting the statistics of return times or waiting times~\cite{martinez2019inferring,skinner2021improved,skinner2021estimating,van2022thermodynamic,harunari2022learn}.

For high-dimensional diffusive systems or Markov jump processes on an ample state space, where the experimental estimate of microscopic forces becomes difficult, the TUR is a valuable tool for quantifying approximately $\sigma$.
Indeed, the TUR is a frugal inequality, being based only on the knowledge of the first two cumulants of any current $J$ integrated over a time $\tau$, i.e~its average $\langle J\rangle$ and its variance $\va(J) = \langle J^2\rangle-\langle J\rangle^2$,
\begin{align}\label{eq:TUR}
     \frac{\sigma }{k_B}\geq 2  \frac{ \langle J\rangle^2}{\va(J) \,\tau}.
\end{align}
However,~\eqref{eq:TUR} might provide a loose bound on $\sigma$. For example, far from equilibrium, kinetic factors~\cite{dit19} often constrain the right-hand side of \eqref{eq:TUR} to values much smaller than  $\sigma$.

Let us focus on a Markov jump process with transition rate $w_{ij}$ from state $i$ to state $j$, in a stationary regime with steady-state probability $\rho_i$. If there is a coupling between equilibrium reservoirs and the system, for each forward transition between two states $i$ and $j$, denoted $(i,j)$, the backward evolution $(j,i)$ is possible.
By defining fluxes $\phi_{ij} = \rho_i w_{ij}$, the mean entropy production rate is written~\cite{maes03_1,seifert2012stochastic,peliti2021stochastic} as
\begin{align}
\label{eq:epr}
     \frac{\sigma }{k_B} &= 
  \sum_{i<j} (\phi_{ij}-\phi_{ji})\ln \frac{\phi_{ij}}{\phi_{ji}} \,.
\end{align}
The system is in equilibrium only if all currents are zero, that is, $\phi_{ij}=\phi_{ji}$ for every pair $\{i,j\}$. Experimentally, fluxes $\phi_{ij}$ are estimated by counting the transitions between states $i$ and $j$ in a sufficiently long time interval $t$. 

The challenge we focus on is evaluating the entropy production rate with experimental data missing the observation of one or more backward transitions---any null flux makes \eqref{eq:epr} inapplicable.  
This situation is encountered in a vast class of idealized mesoscopic systems such as totally asymmetric exclusion processes \cite{derrida2007non}, directed percolation \cite{Takeuchi2007directed}, spontaneous emission in a zero-temperature environment \cite{penocchio2021nonequilibrium}, enzymatic reactions \cite{reuveni2014role}, perfect resetting \cite{mori2023entropy}.
A first solution to this problem involves replacing each unobserved transition $\phi_{ij}$ by a fictitious rate $\phi_{ij} \sim t^{-1}$ scaling with the observation time. 
Intuitively, this corresponds to assuming that no transition was observed during a time $t$ because its flux was barely lower than $t^{-1}$.
A Bayesian approach refines this simple argument, proposing the optimized assumption $\phi_{ij} \simeq (\rho_j/\rho_i) /t$ for null fluxes in unidirectional transitions \cite{zeraati2012entropy}. Hence, it allows us to estimate the entropy production directly from \eqref{eq:epr}.

In this Letter, we put forward an additional estimation of the entropy production rate, not based on \eqref{eq:epr}. Instead, we introduce a lower bound to $\sigma$ (see \eqref{eq:nonlinear_TUR} with~\eqref{eq:main} below) using the exact quantities needed for the estimation of \eqref{eq:epr}, i.e., single fluxes. Notwithstanding that, the new inequality can be tight and give an estimate of $\sigma$ that outperforms \eqref{eq:epr} (and any standard TUR) in regimes lacking data, in which transitions between states may appear experimentally as unidirectional in a trajectory of finite duration $t$. The efficacy of our approach derives from several optimizations related to (i) the short time limit $\tau\to 0$~\cite{man20,otsubo2020estimating}, (ii) the so-called hyperaccurate current~\cite{bus19,busiello2022hyperaccurate,falasco2020unifying,van2020entropy}, and, most importantly, (iii) an uncertainty relation \cite{van2022unified, falasco2020unifying} in which the specific presence of the inverse hyperbolic tangent boosts the lower limit imposed to $\sigma$ by the  TUR \eqref{eq:TUR} (the hyperbolic tangent appears in several derivations of TURs~\cite{hasegawa2019fluctuation,timpanaro2019thermodynamic}).
In the extreme case in which all  observed transitions appear unidirectional, the inequality simplifies to
\color{newcolor}
\begin{align}
\label{eq:main}
 \frac{\sigma}{k_B} 
  &\geq  \kappa \log(\kappa\,t)\,,\qquad\text{for}\;\,\kappa\,t\gg 1,
\end{align}
\color{black}
where the average jumping rate (or dynamical activity, or frenesy~\cite{maes2020frenesy}) $\kappa$ characterizes the degree of agitation of the system. Thus, far from equilibrium, the nondissipative quantity $\kappa$ binds the average amount of dissipation $\sigma$. 
Key to our approach is the assumption that 
the overall rate of all unobserved (reverse) transitions is of the order $\sim t^{-1}$.
At the same time, we make no assumption for the specific reverse rate of each unidirectional transition.
Equation \eqref{eq:main} can also be used if only forward rates are known analytically and backward rates are small and unknown. The chemical reactions described at the end of the paper fall into this scenario.

\paragraph*{\textbf{Results and discussion}}
\paragraph*{Empirical estimation. }
Equation~\eqref{eq:epr} holds for Markov jump processes with transition rates satisfying the local detailed balance condition $w_{ij} / w_{ji}= e^{s_{ij}}$,
where $s_{ij} = - s_{ji}$ is entropy increase in the environment (in units of $k_B=1$, hereafter) when transition $(i,j)$ takes place. 
For these processes, the experimental data we consider are time series of states $(i^{(0)}, i^{(1)},  \ldots, i^{(n)})$ and of the corresponding jumping times $(t^{(1)},t^{(2)},\ldots, t^{(n)} <t)$.  The total residence time $t^R_i$ that a trajectory spends on a state $i$ gives the empirical steady-state distribution $p_i = t^R_i / t$ that approaches the steady-state probability distribution $\rho_i$ for long times $t$.

The estimation of $\sigma$ is based on empirical measurements of fluxes $\phi_{ij}$, which we define starting from the number $n_{ij}=n_{ij}(t)$ of observed transitions $(i,j)$,
\begin{align}\label{eq:flux}
 \phi_{ij} \simeq \dot n_{ij}\equiv \frac{n_{ij}}{t}\,.
\end{align}
If the observation time $t$ is much larger than the largest time scale, i.e. $t\gg \tau_{\rm sys} = (\min_{i,j} \phi_{ij})^{-1}$,
 the empirical flux $\dot n_{ij}$ converges to $\phi_{ij}$ and the estimate of the entropy production rate simply becomes
\begin{align}\label{eq:epr_estimation}
    \sigma \simeq 
    \empir{\sigma}
    \equiv
    \sum_{i<j } (\dot n_{ij}-\dot n_{ji})\ln \frac{\dot n_{ij}}{\dot n_{ji}}.
\end{align}
However, our focus is on systems in which $t < \tau_{\rm sys}$, so that some transitions $(i,j) \in \mathcal{I}$ are probably never observed ($n_{ij}=0$) while their reverse ones $(j,i)$ are ($n_{ji}\ne0$). In this case, the process appears absolutely irreversible~\cite{murashita2014nonequilibrium} and  \eqref{eq:flux} is inapplicable---the estimate \eqref{eq:epr_estimation} would give an infinite entropy production.
We assume that the network remains connected if one removes the transitions belonging to the set $\mathcal{I}$ so that the dynamics stays ergodic.
Note that the case where both $n_{ij}=0$ and $n_{ji}=0$ poses no difficulties. Indeed, one neglects the edge $\{i,j\}$, at the possible price of underestimating $\sigma$. We leave this understood and deal with the residual cases in which only one of the two countings is null.

If in a time $t$ a transition from $i$ to $j$ is not observed, we conclude that the
 typical time scale of the transition is not shorter than $t$, that is $\phi_{ij} \lesssim t^{-1}$. A more quantitative argument~\cite{zeraati2012entropy} suggests ``curing'' the numerical estimates of fluxes by introducing a similar minimal assumption,
\begin{align}\label{eq:regular}
   \dot n_{ij}=
    \begin{cases}
      n_{ij}/t, & \text{if}\ n_{ij}>0 \\
      p_j/(p_i t), & \text{if}\ (i,j) \in \mathcal{I} 
      \\
      0, & \text{otherwise},
    \end{cases}
\end{align}
and uses these regularized estimates of $\phi$'s in \eqref{eq:epr_estimation}.

\paragraph*{Lower bounds to $\sigma$.}

To use TURs, we define currents by counting algebraically transitions between states. For example, for $(i,j)$, an integrated current during a time $\tau$ is just the counting $n_{ij}(\tau) - n_{ji}(\tau) $ during that period.
By linearly combining single transition currents via
antisymmetric  weights $c_{ij}=-c_{ji}$ (stored in a matrix $c$), one may define a generic current $J = J_{c} = \sum_{i<j} c_{ij} [n_{ij}(\tau)  - n_{ji}(\tau)]$.

Among all possible currents, one can choose those giving the best lower bound to $\sigma$, for instance, by machine learning methods~\cite{man20,otsubo2020estimating,kim2020learning}. 
However, Refs.~\cite{busiello2022hyperaccurate,bus19} show that the TUR can be analytically optimized by choosing the hyperaccurate current $J^{\rm hyp}$. In the limit $\tau \to 0$, the coefficients defining $J^{\rm hyp}$ take the simple form~\cite{falasco2020unifying}
\begin{align}
c_{ij}^{\rm hyp}= 
\frac{\phi_{ij}-\phi_{ji}}{\phi_{ij}+\phi_{ji}}
\simeq 
\frac{\dot n_{ij}-\dot n_{ji}}{\dot n_{ij}+\dot n_{ji}}.
\end{align} 
Moreover, the TUR \eqref{eq:TUR} is the tightest when $J$ is integrated over an infinitesimal time $\tau \to 0$~\cite{man20,otsubo2020estimating} and can become equality in the limit $\tau \to 0$ only for overdamped Langevin dynamics. However, for Markovian stationary processes Eq.~(24) in Ref.~\cite{falasco2020unifying} states that 
\begin{align}
\frac{ \langle J_\tau\rangle^2}{\va(J_\tau)\,\tau}
\geq 
\frac{ \langle J_{\mathcal{T}}\rangle^2}{\va(J_{\mathcal{T}})\,\mathcal{T}}
\qquad\text{for}\quad J=J^{\rm hyp},
\end{align}
where $\mathcal{T}\equiv 
M\tau$ is a time span collecting $M$ short steps of duration $\tau$. 
It is therefore useful to exploit the TUR in the short $\tau$ limit and define the short-time precision $\mathfrak{p}(J)$ as
\begin{align}
   \mathfrak{p}(J) \equiv \lim_{\tau \to 0 } \frac{ \langle J\rangle^2}{\va(J)\,\tau}.
\end{align}
For Markov jump processes, for a generic $J$, we have
\begin{align}
\begin{split}
    \langle J\rangle &= \sum_{i < j} c_{ij}( \phi_{ij} - \phi_{ji}) \tau ,\\
    \va(J) &= \sum_{i < j} c_{ij}^2( \phi_{ij} + \phi_{ji}) \tau + O(\tau^2),
\end{split}
\end{align}
and the related TUR in the limit $\tau\to 0$ is
\begin{align}
   \sigma \ge \sigma_{\rm TUR}^J = 2\, \mathfrak{p}(J)\,.
\end{align}
 Focusing on the hyperaccurate current, which is also characterized by $\langle J^{\rm hyp}\rangle=\va(J^{\rm hyp})$, the optimized TUR reduces to $\sigma \ge \sigma_{\rm TUR}^{\rm hyp} = 2 \mathfrak{p}^{\rm hyp}$ (involving the so-called pseudo-entropy~\cite{shiraishi2021optimal}) with
\begin{align}\label{eq:oTUR}
    \mathfrak{p}^{\rm hyp}\equiv  
    \sum_{i < j}
    \frac{( \phi_{ij} - \phi_{ji})^2}{\phi_{ij} + \phi_{ji}}
\end{align}

In the following, we will use an improvement of the TUR. 
We start from the implicit formula derived in \cite{van2022unified},
\begin{align}\label{eq:nonlinear_TUR_implicit}
  \mathfrak{p}(J) \leq \frac{\sigma^2}{4 \kappa {f(\frac{\sigma}{2 \kappa})}^2}
\end{align}
which depends on the frenesy 
\begin{align}\label{eq:frenesy}
\kappa = \sum_{i<j}(\phi_{ij} + \phi_{ji}) \,,
\end{align}
and the function $f$ that is the inverse of $x \tanh x$.
We use the relation $g(x)=x/f(x)$ with $g$ the inverse function of $x \tanh^{-1}\!x$ \cite{nishiyama2022} to turn \eqref{eq:nonlinear_TUR_implicit} into an explicit lower bound on the entropy production 
\begin{align}\label{eq:nonlinear_TUR}
 \sigma \geq  2 \sqrt{\mathfrak{p}^{\rm hyp} \kappa} \,\tanh^{-1}\!\left(\sqrt{\mathfrak{p}^{\rm hyp}/\kappa}\right) .
\end{align}
The inequality \eqref{eq:nonlinear_TUR} reduces to the TUR close to equilibrium, where $\sigma \to 0$, and is tighter than the kinetic uncertainty relation~\cite{dit19} far from equilibrium where $\tanh (\sigma/2\sqrt{\mathfrak{p}^{\rm hyp} \kappa }) \to 1$, namely $\kappa  \geq \mathfrak{p}^{\rm hyp}$.

\color{newcolor}
Empirically, for $\kappa\,t\gg 1$, we can approximate the optimized precision \eqref{eq:oTUR} and the frenesy \eqref{eq:frenesy} with
\begin{align}
\label{eq:phyp-e}
\empir{\mathfrak{p}}^{\rm hyp}
   & =
    \sum_{i < j}
    \frac{( \dot n_{ij} - \dot n_{ji})^2}{\dot n_{ij} + \dot n_{ji}}\,,\\
\label{eq:frenesy-e}
\empir{\kappa}
& = 
\sum_{i<j}(\dot n_{ij} + \dot n_{ji})\,.
\end{align}
Neither of them does require the regularization \eqref{eq:regular}. 
However, the (positive) argument of the $\tanh^{-1}$ function in \eqref{eq:nonlinear_TUR} needs to be strictly lower than $1$, i.e.~$\empir{\mathfrak{p}}^{\rm hyp}<\empir{\kappa}$.

With $\empir{\mathfrak{p}}^{\rm hyp}$ there arises a problem when one measures irreversible transitions in all edges, $\min(\dot n_{ij},\dot n_{ji})=0$ for every $i$ and $j$: in that case, one can see that $\empir{\mathfrak{p}}^{\rm hyp} = \empir{\kappa}$.
To fix the divergence of $\tanh^{-1}$ that it would lead to, we use an assumption milder than the requirement of reversibility for all transitions used for \eqref{eq:regular}.

If $\empir{\mathfrak{p}}^{\rm hyp} = \empir{\kappa}$, we assume that the observation time $t$ is barely larger than the typical time needed to have any reverse transition.
By denoting a tiny rate of any unobserved transition $(i,j)\in {\cal I}$ as $\epsilon_{ij}$, the ratio of the precision of the hyperaccurate current over the frenesy becomes
\begin{align}\label{eq:ratio}
    \frac{\mathfrak{p}^{\rm hyp}}{ \kappa}
    &\simeq 
    \frac{1}{
     \sum_{(i,j)\in {\cal I}}(\dot n_{ji} +\epsilon_{ij})
    }
    \sum_{(i,j)\in {\cal I}}
    \frac{( \dot n_{ji} - \epsilon_{ij})^2}{\dot n_{ji}+\epsilon_{ij}} 
    \nonumber\\
    &\simeq 
    \frac{
    \empir{\kappa} - 3
     \sum_{(i,j)\in {\cal I}}\epsilon_{ij} }{
    \empir{\kappa} + 
     \sum_{(i,j)\in {\cal I}}\epsilon_{ij}
    }
    \nonumber\\
    &\simeq 
    1 - \frac{4}{\empir{\kappa}} \sum_{(i,j)\in {\cal I}}\epsilon_{ij}\nonumber\\
    &= 1 - \frac 4 {\empir{\kappa}\,t} 
\,\simeq 1 - \frac 4 {\kappa\,t} ,
\end{align}
where we have set $\sum_{(i,j)\in {\cal I}}\epsilon_{ij} = 1/t$ according to our assumption and we replaced $\empir{\kappa}$ with $\kappa$ because they give the same bound to leading order in $t$.

Hence, in an experiment measuring only irreversible transitions  for $t\gg 1/\kappa$, \eqref{eq:nonlinear_TUR} with \eqref{eq:ratio} still provide a lower bound on the entropy production rate:
\color{newcolor}
\begin{align}
\label{eq:main2}
 \frac{\sigma}{k_B } 
  &\geq  2 \kappa \tanh^{-1}\left(\sqrt{1-\frac {4} {\kappa\,t} }\right) 
\end{align}
By expanding $\tanh^{-1}$ to the leading order,~\eqref{eq:main2} simplifies to the lower bound anticipated in~\eqref{eq:main}, which is based simply on evaluating the frenesy $\kappa$ empirically with~\eqref{eq:frenesy-e}.

\color{black}

\begin{figure}[!t]
\centering
\includegraphics[width=0.98\linewidth]{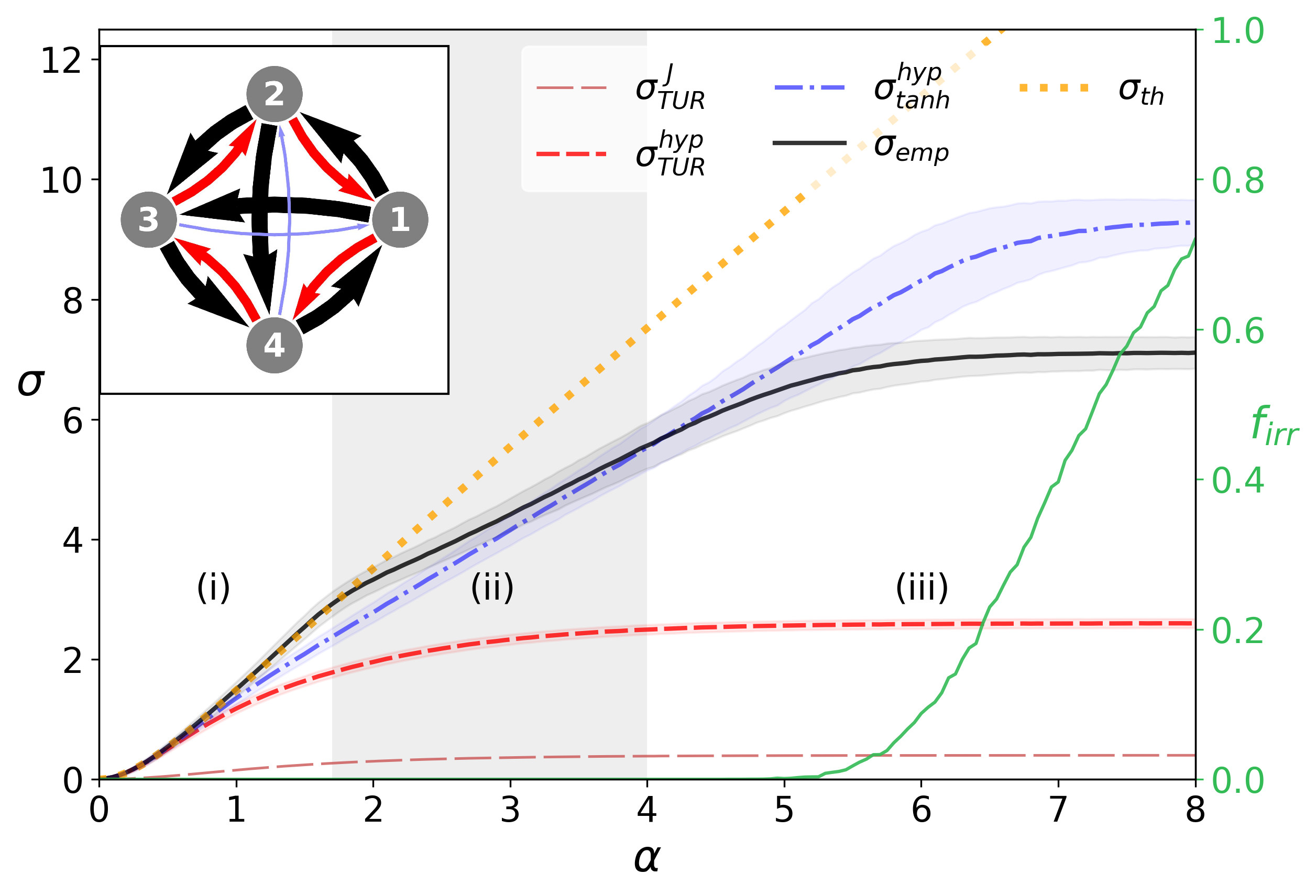}

\caption{{\bf Estimation of the entropy production in a 4-state model.} For the 4-states model (inset and description in the text), estimates of the entropy production rate and theoretical value $\sigma_{\rm th}$ (left axis), and the fraction of trajectories displaying only irreversible transitions (green curve, right axis) as a function of the nonequilibrium strength $\alpha$.
The sampling time is $t=10^3$, and bands show one standard deviation variability over trajectories. Highlighted regions: (i) the empirical estimate works well while lower bounds progressively depart from $\sigma_{\rm th}$; (ii) $\empir{\sigma}$ deviates from $\sigma_{\rm th}$ but remains the best estimator; (iii) the new lower bound $\sigma_{\tanh}^{\rm hyp}$ is the best estimator.
}
\label{fig:4states}
\end{figure}

\paragraph*{Examples.}

Let us illustrate the performance of the various estimators of the entropy production rate with the four-state nonequilibrium model sketched in Fig.~\ref{fig:4states}. Some transitions (black arrows) have a constant rate $w_{ij}=1$ (in dimensionless units). Other transition rates depend on a nonequilibrium strength $\alpha$, with $\alpha=0$ representing equilibrium: $w_{ij}=e^{-\alpha}$ (red arrows), and $w_{ij}=e^{-3 \alpha}$ (thin blue arrows). The latter class is the first to go undetected for sufficiently large $\alpha$, which causes the empirical estimate $\empir{\sigma}$ to start deviating from the theoretical value ($\alpha \gtrsim 1.7$ in Fig.~\ref{fig:4states}). However, $\empir{\sigma}$ remains the best option for evaluating $\sigma$ up to a value $\alpha \approx 4$, where the lower bound $\sigma_{\tanh}^{\rm hyp}$ takes over as the best $\sigma$ estimator far from equilibrium. Interestingly, for $\alpha \gtrsim 4$, the probability $f_{\rm irr}$ to measure a trajectory with only irreversible transitions (green curve in Fig.~\ref{fig:4states}) is still small. Hence, it is the full scheme with \eqref{eq:nonlinear_TUR} and the extreme case~\eqref{eq:main} that provides a good estimate of $\sigma$.
For this setup, the TUR optimized with the hyperaccurate current is only useful close to equilibrium but is never the best option. For comparison with the hyperaccurate version, in Fig.~\ref{fig:4states}, we also show the loose lower bound offered by the TUR for the current $J$ defined on the single edge $\{i=1, j=2\}$.

\begin{figure}[!t]
\centering
\includegraphics[width=0.99\linewidth]{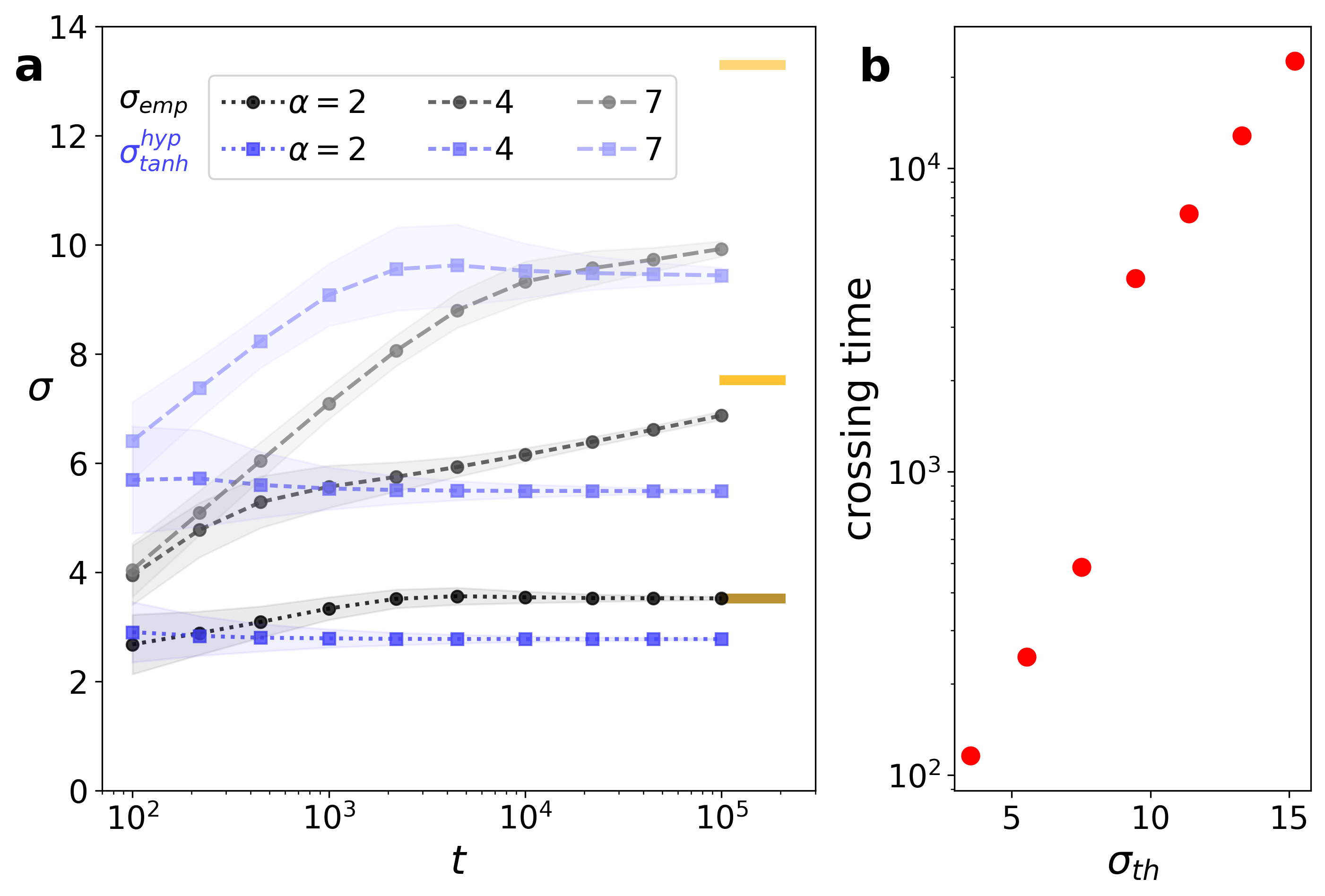}
\caption{ {\bf Estimation of the entropy production as a function of the sampling time.}
{\bf a.} For three values of the nonequilibrium parameter $\alpha$ (see legend) in the model of Fig.~\ref{fig:4states}, scaling with the sampling time $t$ of the estimators $\empir{\sigma}$ and $\sigma_{\tanh}^{\rm hyp}$, and theoretical values (horizontal thick lines). Bands show one standard deviation variability over trajectories.
{\bf b.} Time when $\empir{\sigma}$ becomes larger than $\sigma_{\tanh}^{\rm hyp}$, as a function of $\sigma_{\rm th}$.
}
\label{fig:4statestime}
\end{figure}

To appreciate the influence of the trajectory duration on the estimators,
in Fig.~\ref{fig:4statestime}a, we plot the scaling with the sampling time $t$ of $\empir{\sigma}$ and $\sigma_{\tanh}^{\rm hyp}$. In this example, the farther one goes from equilibrium, the longer $\sigma_{\tanh}^{\rm hyp}$ remains the better estimator. This aspect can be crucial if, for experimental limitations, one is restricted to a finite $t$. It is also emphasized in Fig.~\ref{fig:4statestime}b, where we plot the time at which $\empir{\sigma}$ becomes larger than $\sigma_{\tanh}^{\rm hyp}$, as a function of the true dissipation rate $\sigma_{\rm th}$. Again, it shows that in more dissipative regimes, one can rely on $\sigma_{\tanh}^{\rm hyp}$ if trajectories are too short to obtain a good estimate of $\sigma_{\rm th}$ with $\empir{\sigma}$.

\begin{figure}[!t]
\centering
\includegraphics[width=\linewidth]{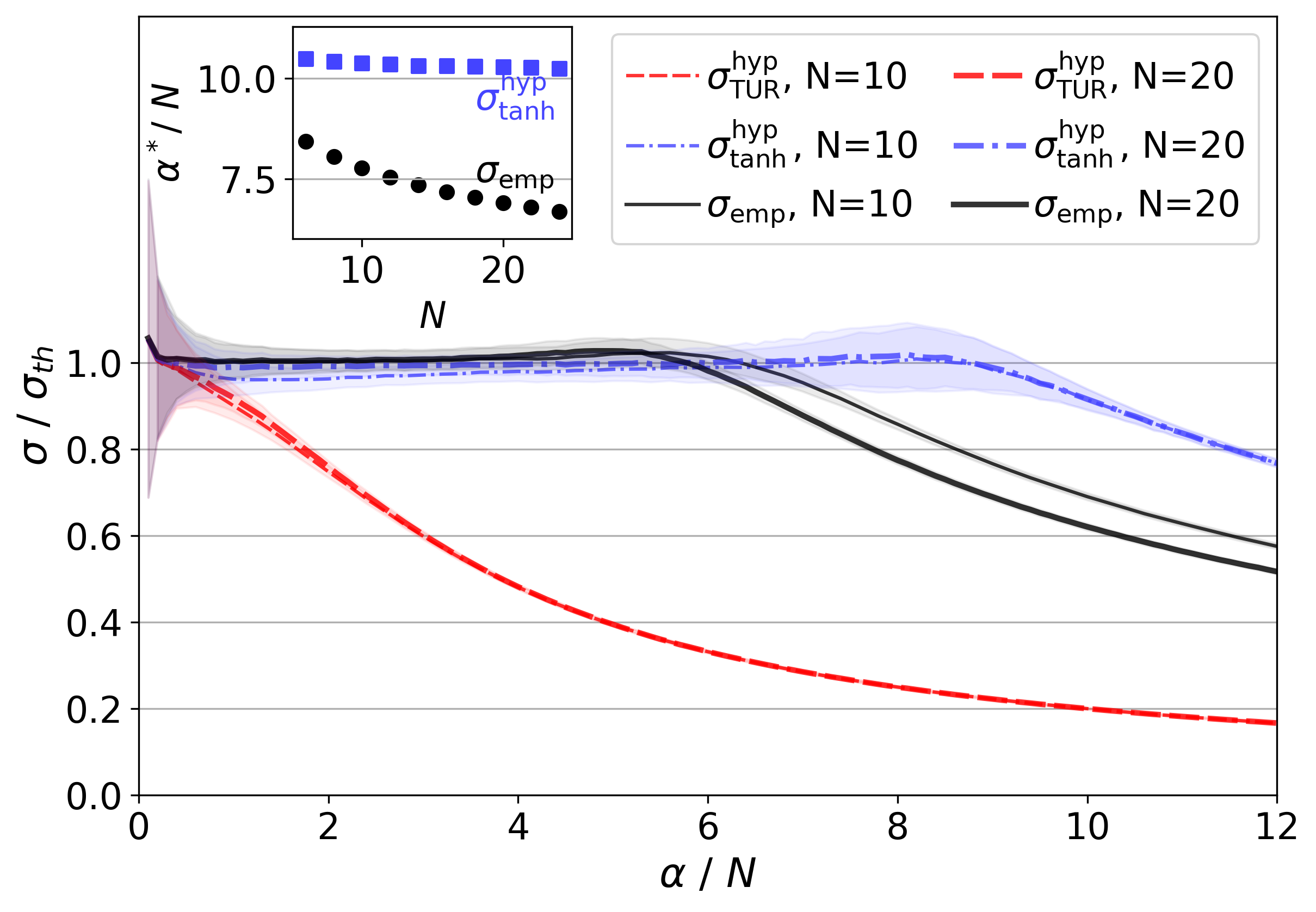}
\caption{{\bf Estimation of the entropy production for a ring model.} For the ring model described in the text, we show various estimates of the entropy production rate divided by the theoretical value, for $t=10^4$, as a function of the nonequilibrium strength $\alpha/N$, for ring lengths $N=10$ and $N=20$. The inset shows $\alpha^*/N$.
For $\alpha\to 0$,  the standard deviation of $\sigma/\sigma_{\rm th}$ (shaded bands) are amplified because also $\sigma_{\rm th}\to 0$.}

\label{fig:ring}
\end{figure}

The second example shows how the new lower bound may scale favorably with the system size compared to the other $\sigma$ estimators.
We study a periodic ring of $N$ states with local energy $u_i = -\cos(2\pi i/N)$ and transitions rates $w_{i,i+1}=1$, $w_{i,i-1}=\exp[-\alpha/N + u_i-u_{i-1}]$ (temperature is $T=1$). 
Fig.~\ref{fig:ring} shows the ratios of $\sigma^{\rm hyp}_{\rm TUR}$, $\sigma^{\rm hyp}_{\tanh}$, and $\empir{\sigma}$ over the theoretical value $\sigma_{\rm th}$, as a function of the nonequilibrium force $\alpha$, both for a ring with $N=10$ and for a longer ring with $N=20$ states. For each $N$, we see that $\sigma^{\rm hyp}_{\tanh}$ far from equilibrium is the best estimator of the entropy production rate. Furthermore, it also appears to be the estimator that scales better by increasing the system size: its plateau at $\sigma/\sigma_{\rm th}\simeq 1$ scales linearly with $N$ (the inset of Fig.~\ref{fig:ring} shows the $\alpha^*$ value where $\sigma/\sigma_{\rm th}$ drops below the arbitrary threshold $0.9$, divided by $N$, as a function of $N$), while this is not the case for the empirical estimate $\empir{\sigma}$.
These results propose $\sigma^{\rm hyp}_{\tanh}$ as a valuable resource in a continuous limit to deterministic, macroscopic conditions. 
Next, we explore this possibility.

\paragraph*{Deterministic limit.}
Our approach extends to deterministic dynamics resulting from the macroscopic limit of underlying Markov jump processes for many interacting particles such as driven or active gasses~\cite{Bertini2015Jun,solon2013revisiting}, diffusive and reacting particles \cite{gaveau1999variational,falasco2018information}, mean-field Ising and Potts models \cite{meibohm2022finite,herpich2018collective}, and charges in electronic circuits \cite{Freitas2021}. In these models, the state's label $i$ is a vector with entries that indicate the number of particles of a given type. The system becomes deterministic when the typical number of particles goes to infinity, controlled by a parameter, such as a system size $V \to \infty$. In this limit, it is customary to introduce continuous states $x=i/V$~\cite{vankamp1992stochastic}, e.g., a vector of concentrations, such that $w_{ij} = V \omega_r(x)$ where $\pm r$  labels the transitions to (or from) the states infinitesimally close to $x$. Such transition rate scaling is equivalent to measuring events in a macroscopic time $\tilde t \equiv t V$. 
The probability $p_i \sim \delta(x-x^*)$ peaks around the most probable state $\langle x\rangle\equiv x^*$. The entropy production rate \eqref{eq:epr_estimation} becomes extensive in $V$ \cite{herpich2020stochastic}, and its density takes the deterministic value
\begin{align}
  \tilde{\sigma} \equiv \frac{ \sigma}{V} =  \sum_{r>0} [\omega_r(x^*)-\omega_{-r}(x^*)] \ln \frac{\omega_r(x^*)}{\omega_{-r}(x^*)},
  \label{tilde-emp}
\end{align}
with $\omega_{\pm r}(x^*)$ the macroscopic fluxes.

\begin{figure}[t!]
\centering
\includegraphics[width=0.98\linewidth]{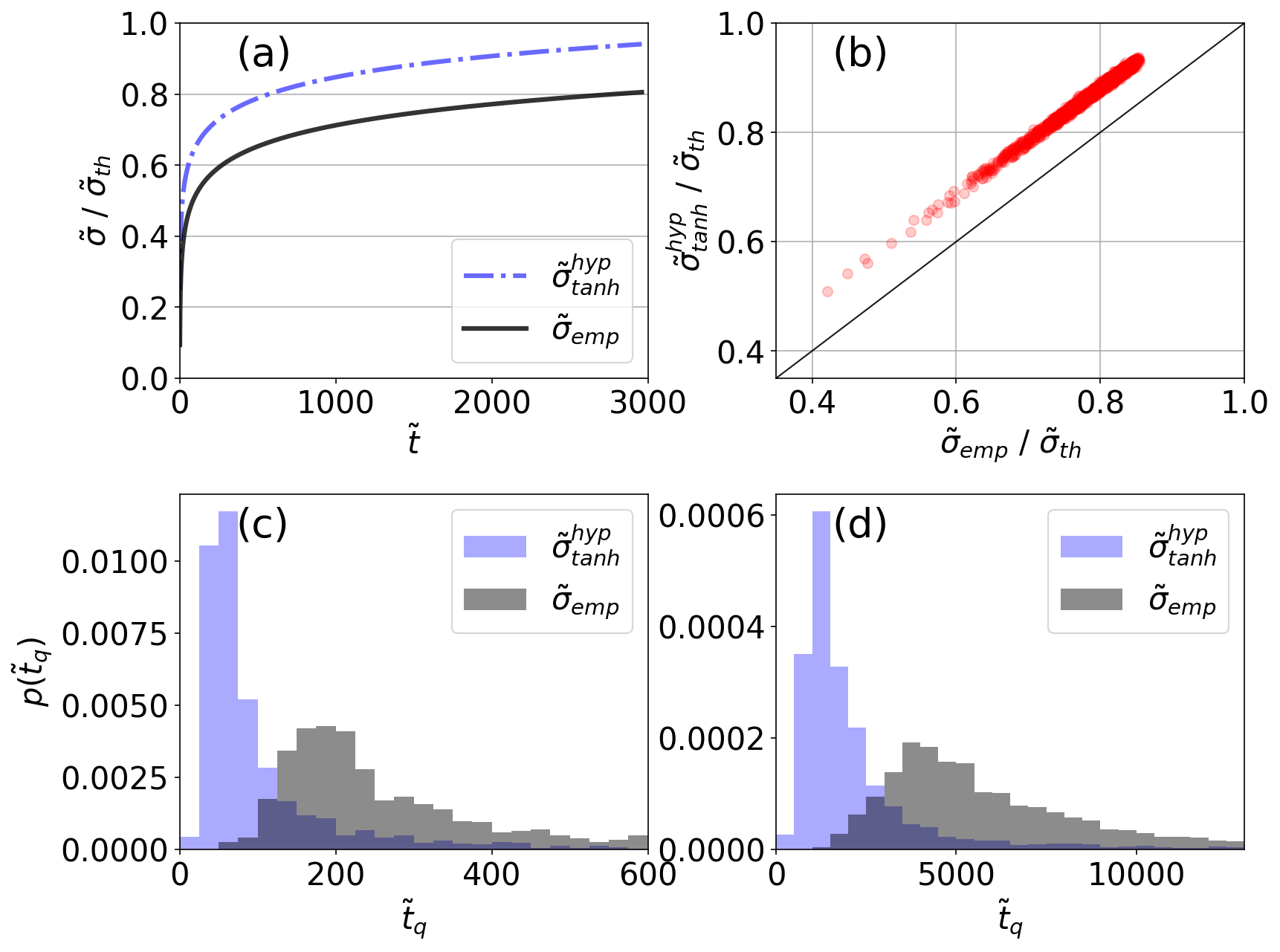}
\caption{{\bf Estimation of the entropy production in a deterministic chemical reaction network.} \textbf{a.} Estimates \eqref{tilde-emp} and \eqref{tilde-hyp} normalized to the true $\sigma_{\rm th}$ as a function of macroscopic time $\tilde t$ up to the inverse of the maximum flux, for the chemical reaction network~\eqref{eq:CRN} with forward rate constants $k_{+r}= \{5,2,1,0.2\}$ and backward $k_{-r} =10^{-5}$. \textbf{b.} With $2000$ random realizations of the chemical reaction network (uniformly sampled rate constants $k_{+r}\in [10^{-2},10^2]$, and $k_{-r} =10^{-5}$), estimate \eqref{tilde-emp} vs.~\eqref{tilde-hyp}, both normalized to the true $\tilde \sigma_\text{th}$, 
for a fixed time $\tilde t=0.1/ \max(\omega_r(x^*))$; 
\textbf{c.}  probability distribution of $\tilde t_q$ for $q=0.6$, and \textbf{d}  $q=0.8$.
}
\label{fig:chem}
\end{figure}

We are interested in the case where backward transitions ($r<0$) are practically not observable, i.e., the fluxes $\omega_{-|r|}(x^*)$ are negligibly small compared to the experimental errors.
Since $\kappa$ is also extensive in $V$, we define the frenesy density $\tilde \kappa \equiv \kappa/V = \sum_{r} \omega_r(x^*)$ and apply \eqref{eq:main}  as
\color{newcolor}
\begin{align}
  \tilde \sigma \geq   \tilde \kappa \log(\tilde\kappa \,\tilde t).
  \label{tilde-hyp}
\end{align}
\color{black}
The formula above holds for $\tilde t \ll 1/ \max(\omega_{-|r|}(x^*))$, which is the typical time when backward fluxes become sizable.

We compare \eqref{tilde-emp} and \eqref{tilde-hyp} for systems of chemical reactions with mass action kinetics, i.e., each flux $\omega_{\pm r}(x^*)$ is given by the product of reactant concentrations times the rate constant $k_{\pm r}$ \cite{rao2016nonequilibrium}. In particular, we take the following model of two chemical species, X and Y, with uni- and bimolecular reactions,
\begin{align}\begingroup
\begin{aligned}
& \ce{ $\emptyset$  <=>[$k_{+1}$][$k_{-1}$]  X } \qquad
\ce{2X  <=>[$k_{+2}$][$k_{-2}$] X +Y } \\
& \ce{X  <=>[$k_{+3}$][$k_{-3}$]  Y } \qquad
\ce{Y  <=>[$k_{+4}$][$k_{-4}$] \emptyset .}
\end{aligned}
\endgroup
  \label{eq:CRN}
\end{align}
Fig.~\ref{fig:chem}a shows, for a specific set of rate constants, that the new bound \eqref{tilde-hyp} outperforms the empirical estimation \eqref{tilde-emp} for all times when the dynamics appears absolutely irreversible. 
This occurs for all randomly drawn values of the rate constants (Fig.~\ref{fig:chem}b). 
Additionally, we plot the histogram of times $\tilde t_q$ at which \eqref{tilde-hyp} and \eqref{tilde-emp} reach the fraction $q<1$ of the theoretical entropy production rate, for $q=0.6$ in Fig.~\ref{fig:chem}c and $q=0.8$. Fig.~\ref{fig:chem}d. Times resulting from \eqref{tilde-hyp} are significantly shorter than those from the empirical measure \eqref{tilde-emp}.

The application of \eqref{eq:main} is possible either when the forward rates are analytically known (as in the example in Fig.~\ref{fig:chem}) or when they can be experimentally reconstructed. Direct measurement of chemical fluxes is a challenging task. One case in which they are measurable with high precision is photochemical reactions in which the emission of photons at a specific known frequency signals the reaction events (see \cite{penocchio2021nonequilibrium} and references therein). In general, one can measure reliably only concentrations, from which one can extract reaction fluxes only in specific networks (and if the reaction constants are known). We note, however, that the same method exemplified with chemical reactions can be used, e.g., for electronic circuits, where counting electron fluxes between resistive elements is much easier.

\paragraph*{\textbf{Conclusion} }
In summary, the lower bound \eqref{eq:nonlinear_TUR}, turning into~\eqref{eq:main} for systems that appear absolutely irreversible, is more effective than the direct estimate \eqref{eq:epr_estimation} with~\eqref{eq:regular} in cases of lacking data.
Knowledge of the macroscopic fluxes is enough to apply our new formula \eqref{eq:main}, which outperforms the direct estimation in strongly irreversible systems where all backward fluxes are undetectable.
Thus, we provide a new effective tool to estimate the dissipation in biological and artificial systems, whose performances are limited by energetic constraints~\cite{yang2021physical,Gao2021principles,freitas2022reliability}.

\paragraph*{\bf Acknowledgments}
We thank the anonymous reviewers for their insightful comments. Funding from the research grant BAIE\_BIRD2021\_01 of Universit\`a di Padova is gratefully acknowledged.
GF is funded by the European Union -- NextGenerationEU -- and by the program STARS@UNIPD with project ``ThermoComplex''.

\end{document}